# Birth and Early Growth of Entanglement by $sd$ Exchange with Gate-Voltage-Controllable Destiny

Son-Hsien Chen (陳松賢)[1, *]

[1]*Department of Applied Physics and Chemistry, University of Taipei, Taipei 10048, Taiwan*
(Dated: October 24, 2023)

We investigate bipartite entanglement between two distant parties, $A$ and $B$, comprising local magnetic impurities (or qudits) induced by the quench through $sd$ exchange in a field-effect-transistor geometry. A wave-function-based time-dependent formalism is employed by including non-dissipative responses that allow for the control of entanglement via gate voltages. Our study focuses on the birth and early growth of entanglement, by introducing environment support states that render site- and layer-resolved logarithmic negativity (LN) and mutual information (MI). In the minimal set, where party $A$ ($B$) consists of a qubit, we identify entanglement sudden deaths (ESDs), which are explained by a visualization picture analyzing the density matrix. Vibrating electron currents facilitate the birth of entanglement, while they are not required for its growth and subsistence. The LN emerges near the edge layers in $A$ and $B$, while MI shows up outside these two parties within the spacing layer. The MI is born earlier than the LN. When a gate voltage large enough to disjoint part of the system is applied within the spacing region, it partially suppresses the entanglement, quantified by the LN. This suppression does not appear immediately after the presence of the disjoint voltage. Applying this disjoint voltage to the site(s) hosting the qudit(s) helps prevent the site- and layer-resolved LN from encountering ESDs. The local impurities in parties $A$ and $B$ are initially of opposite spin directions in an unentangled state, as can be prepared by two of our proposed protocols. However, the features described above do not depend on the chosen protocols.

## I. INTRODUCTION

Entanglement, signifying that a total wave function can not be expressed separately in the tensor product form of the corresponding subsystem wave functions, enables distinguishing quantum information processing and quantum bit (qubit) functionality from classical counterparts. As a pure quantum feature, entanglement underpins many cutting-edge technologies including gravitational wave detections [1, 2], quantum cryptography [3–6], and quantum computation [7–12]. Entanglement has shown to be helpful in comprehending, analyzing, and exploring fundamentals in physics, gauge physics and spin dynamics for example. Recently, with the aid of a state-of-the-art trapped-ion quantum computer [13], nonequilibrium phenomena within a lattice gauge theory were formulated [14], permitting the observation of a dynamic topological quantum phase transition and the extraction of Rényi entropies [15], fidelities, and reduced density matrices (DMs). In spintronics devices, vanished dynamic phases and discretized (into 0 or $\pi$) geometric phases at maximum entanglement were identified [16] by injecting an entangled spin pair into a quantum square ring possessing non-Abelian gauge [17]. It has also been discovered [18, 19] that the entanglement plays a crucial role for the applicability of the time-honored Landau-Lifshitz-Gilbert (LLG). Since the entanglement shrinks the spin polarization (or magnetization) magnitude which is a conserved quantity in the LLG equation, a regime where entanglement diminishes will admit the validity of this equation. One such regime is recognized in the Kondo lattice model with small enough $sd$ exchange (i.e., exchange between itinerant electron and local spins) coupling [18].

Experimentally, inelastic neutron scattering [20–22] has been employed to probe entanglement in spin chains by examining the quantum Fisher information. Moreover, recent proposals based on X-ray scattering [23, 24] have made entanglement detection in macroscopic solids accessible. Also has been demonstrated is that the dynamics of the von Neumann entanglement entropy $\mathcal{S}$ [25], namely, the Rényi entropy [15] $\mathcal{S}_r$ of index one $r = 1$, in isolated systems renders a buildup similar to the maximization of thermodynamic entropy, which, in the long-time limit after thermalization, follows the volume-scaling law, i.e., is proportional to the system size [26, 27].

The initial growth of Rényi-type entropy induced by the quench (or sudden switch) exhibits interesting behavior. With preparing separable (or unentangled) tensor-product states, from, for instance, the system partitioned into two independent halves before the sudden joint coupling occurs, the quench evokes not only bipartite entropy but also elicits local-observable spreading. Specifically, the entanglement of a local observable decreases its magnitude, leading to the loss of its corresponding purity. This purity-loss information, characterized by the updated wave front, propagates or spreads in the surrounding environment. In ergodic systems, the spreading is governed by some maximum velocity, defining the so-called light cone, and it was found that the early developed entropy $\mathcal{S}$ grows linearly in time [28–31]. By contrast, in localized systems such as the Anderson impurity model, $\mathcal{S}$ emerges logarithmically [32].

Nonetheless, regarding the bipartite entanglement,

---

[*] sonhsien@utaipei.edu.tw

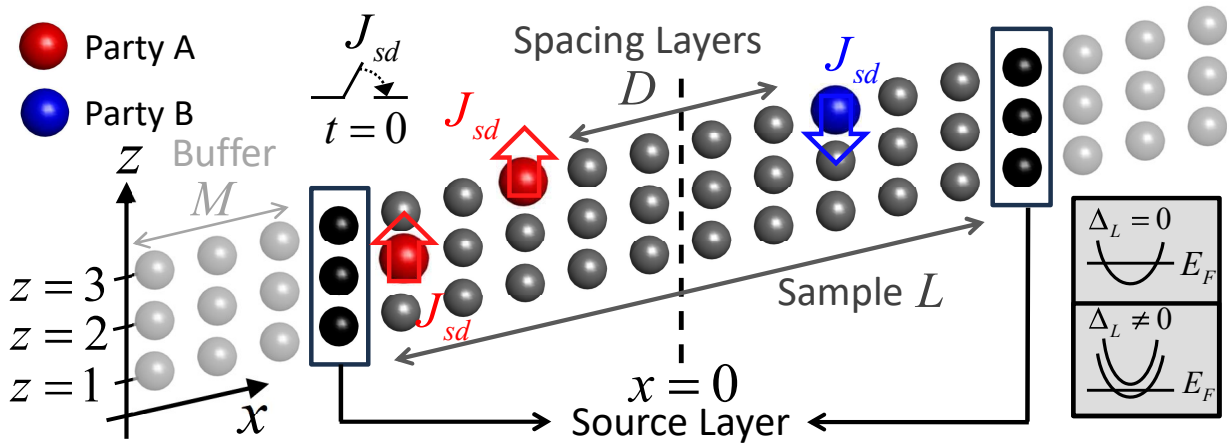

FIG. 1. Studied two-terminal three-atom-wide device geometry consisting of three subsystems, parties $A$ represented by red arrows, $B$ by the blue arrow, and $C$ by the grayscale dots. Local magnetic impurities, functioning as qudits, form parties $A$ and $B$ of opposite spin directions as indicated by the arrows. Conducting electrons at Fermi energy $E_F$ in party $C$, as the mediating environment, grow entanglement between parties $A$ and $B$ via the quench by $sd$ exchange $J_{sd}$ switched on at time $t = 0$. Party $C$ contains three portions, $L$ sample layers hosting impurities, two source layers accounting for the initial conditions of the leads, and $M$ buffer layers to simulate open boundary conditions. We refer the sample-source layer as the active region herein. Also, $A$ and $B$ are separated with $D$ spacing layers centered at $x = 0$. Top (bottom) panel in the lower right corner represents the band-structure schematics for preparing initial impurity spin states in the protocol without (with) necessity of the local Zeeman splitting $\Delta_L$; these schematics will be used to indicate which protocol is used throughout the figures herein.

*only for pure states* (quantified by $\mathcal{S}_{r=2}$), the Rényi-type entropy and its variant such as the mutual information [33, 34] (MI) $\mathcal{M}$ are faithful, i.e., nonzero $\mathcal{S}$ and/or $\mathcal{M}$ implies nonzero entanglement and vice versa. In systems of mixed states, by contrast, there can be classical correlations that originate from the mixture of, or a weighted average over, pure states; the $\mathcal{S}$ and $\mathcal{M}$ in general encode both classical and quantum correlations (entanglements) [35–38]. However, one measure that exists for distilling entanglement from mixed states is the logarithmic negativity [39] (LN) $\mathcal{L}$. While $\mathcal{L} > 0$ in general is a sufficient but not a necessary condition for entanglement, in the special case of two-qubit entanglement, $\mathcal{L}$ and another measure called concurrence [40–42] $\mathcal{C}$ are found to be faithful [43, 44]. A comparison between different entanglement monotones for witnessing entanglement is examined in the recent paper, Ref. [45].

The evolution of negativity has also drawn attention in more recent studies [46–48]. For example, consider a one-dimensional (1D) free fermion chain cut into halves before the quench by a defect acting as a sudden switch for the joint hopping. The quench induces LN that initially grows logarithmically in time [46], while eventually, in the regime where the volume-scaling law applies, saturates at some value characterized by the Rényi mutual information of index half. Contrarily, in the case of a quench induced by a dissipative impurity that absorbs the fermions [48], the negativity does not conform to the saturation depicted by the half-indexed Rényi mutual information. In addition, the concurrence was employed in maximally entangled mixed states [49] to identify a peculiar phenomenon known as entanglement sudden deaths (ESDs). In ESDs, entanglement only persists for a finite time interval before being extinguished by the environment [49–51]. Multiple sequential ESDs often emerge together, and so do the corresponding rebirths.

Only few studies, however, have addressed *distant* bipartite entanglement in *finite-width* open spintronics devices connected to leads with finite level broadening. Incorporating the Fermi-energy or Fermi-Dirac distribution into the formulation of entanglement is rarely seen. Moreover, a transparent picture to comprehend, predict, and prevent ESDs, is especially desirable for solid-state systems.

In this paper, we investigate entanglement between two distant parties, $A$ and $B$, consisting of local spins (or magnetic impurities) mediated by the environment, party $C$ of conduction electrons, in a geometry resembling nanoelectronics field-effect transistors (FETs) [52, 53] with a few spacing layers shown in Fig. 1. The local impurity carries a spin of $S \times \hbar$ acting as a qudit of dimension $d$. For example, $S = 1/2$ represents a qubit, and $S = 1$ represents a qutrit. Specific quench is considered by turning on the $sd$ exchange coupling $J_{sd}$ at time $t = 0$. We adopt initial separable states of a spin-antiparallel configuration, meaning that all prepared spins in party $A$ ($B$) are oriented along the $z$ ($-z$) direction before the quench. This configuration is known to yield significant entanglement as in antiferromagnets [22, 23, 54, 55], and it ensures that the torques exserting on parties $A$ and $B$ are purely quantum-mechanical [56, 57]. We implement the wave-function-based time-dependent formalism developed in Ref. [58] to construct DMs and nonequilibrium lesser Green functions (GFs) for computing the concurrence $\mathcal{C}$, mutual information $\mathcal{M}$, logarithmic negativity $\mathcal{L}$, as well as cross-sectional charge currents. Two protocols for preparing the initial states are offered, one without the need for local Zeeman-splitting $\Delta_L$ applied to the impurities (top panel schematics in the lower right corner of Fig. 1) and the other with the need for $\Delta_L$ (bottom panel schematics of Fig. 1). However, our main findings, which we briefly summarize below, do not depend on these protocols.

Figure 2 exhibits two-qubit entanglement with ESDs identified in Fig. 2(a), and analysis performed to render a visualization picture 2(d) capturing the entanglement fragility from mediating environments. Contrarily, beyond the minimal set of two qubits, the entanglement is robust against ESDs. The negativity yields an overall decaying with $\Delta_L$ in Fig. 3, occurring at a long time after the transient contribution has completely diminished. While, in general, $\mathcal{L}$ decays with the distance $D$ between parties $A$ and $B$, oscillatory behavior resembling the Ruderman-Kittel-Kasuya-Yosida (RKKY) [59–61] exchange can also appear especially when the impurity resides at the bulk site in Fig. 4. Figure 5 illustrates the growth of $\mathcal{L}$ and $\mathcal{M}$, which emerge at some *finite* (*non-zero*) time after the quench. In party $C$, the quench drives vibrating currents to establish $A$-$B$ entanglement (namely, entanglement between $A$ and $B$), demonstrated in Fig. 6. The currents vibrate with respect to the center of the spacing region, even before the onset of entanglement. Inspecting layer-resolved LN $\mathcal{L}_l$ and MI $\mathcal{M}_l$ reveals that LN is born near the edge layers in party $A$ and party $B$, whereas MI is born outside these two parties in the middle of the spacing region (see Fig. 7). Applying a large disjoint gate voltage $eV_g$ at time $t_{g,on}$ within the spacing region results in partial suppression of the entanglement LN, in the case when $eV_g$ is turned on right after the birth Fig. 8 as well as in the case when $eV_g$ is on during the growth Fig. 9. In the limit of $t_{g,on} \to 0^+$, no LN is found, layer-resolved $\mathcal{M}_l$ vanishes, while $\mathcal{M}$ remains growing, attributing this growth to the classical correlations rather than entanglements. We show that $eV_g$ can serve as a remedy for ESDs when applied at one (or both) of the impurity site, comparing Fig. 10 with Fig. 2(a).

The paper is organized as follows. We first describe the studied device geometry and modeling in Sec. II A, where the time-dependent formalism for constructing the DM is also introduced. We briefly review the employed entanglement monotones in Sec. II B. Starting with the minimal set of two qubits, we inspect ESDs in Sec. III A. Section III B elucidates the birth and growth of the entanglement beyond the minimal set, while Sec. III C unveils how entanglement can be adjusted by the gate voltage that disconnects or isolates part of the system. The results are summarized in Sec. IV.

## II. FORMALISM

In this section, we define our device geometry and introduce the time-dependent formalism for evaluating entanglement and charge currents. In modeling nanoelectronics dynamics, the approaches can be categorized into two methods, the partition method [62–66] and the partition-free method [67–71]. The former assumes that, before time $t = 0$, the system is divided into decoupled subsystems in their own thermal-equilibrium. The later considers that the whole open system with a sample connecting to thermal reservoirs or leads is already in equilibrium at time $t = 0$. Although the two methods are equivalent [58, 72], the assumption in the partition-free formalism is often the realistic situation encountered, and practically it bypasses the computation needed for simulating the system to reach the desired preparation of global equilibrium. In the partition-free method, one can formulate the dynamics within the framework of GFs [73–75] or of wave functions [58, 76, 77]. Another approach known as the time-dependent DM renormalization group [78–83], truncating the number of basis to arrive at a matrix product state, is efficient particulary for low-entangled 1D systems. Among the above formalisms, we choose to adopt the wave-function-based formalism [58], due to its direct and concise relations to physical quantities; it also enables encapsulating FET geometry via semi-infinite leads of Fermi-Dirac distributions as described below.

### A. Device geometry and time-dependent density-matrix formalism

The device of interest comprises three parties. Party $A$ ($B$) consists of $N_A$ ($N_B$) local impurities of magnetic moments, each carrying an initial spin of $S \times \hbar$ in $+z$ ($-z$) direction, which acts as a qudit with dimension $d = 2S + 1$. Party $C$ consists of conducting electrons in an infinite long, two-terminal wire with a finite width, such as the case we focus on here with three atomic sites ($z = 1, 2,$ and $3$ in Fig. 1). The investigated entanglement is bipartite between parties $A$ (indicated by the red arrows in Fig. 1) and $B$ (blue arrows in Fig. 1), while party $C$ (sites in grayscale, Fig. 1) serves as a medium or environment to foster the entanglement. The separation distance between parties $A$ and $B$ is determined by the number of $D$ spacing layers, centered at $x = 0$. Party $C$ is divided into three portions, the sample of $L$ layers, two source layers, and $M$ buffer layers. The sample segment, excluding the spacing layers, indicates the areas allowed to host impurities. The source layers specify the sample boundary and account for the initial conditions (ICs) of the left and right leads. The buffer layers simulate open boundary conditions and are elongated by detecting the wave function prorogations to avoid any reflections back to the source and sample regions, i.e., $M$ in general is a time-dependent function $M = M(t)$. Parties $A$ and

B are coupled with the surrounding electrons in Party C via the sd exchange (or Kondo exchange) $J_{sd}$, which is switched on suddenly (leading to the quench) at time $t = 0$. Before $t = 0$, parties A and B are unentangled in the anti-parallel spin configuration, as indicated by the arrow directions in Fig. 1. To prepare these initial unentangled states, we provide two scenarios or protocols, one with $\Delta_L = 0$ and the other $\Delta_L \neq 0$. In the $\Delta_L = 0$ protocol, no local Zeeman splitting $\Delta_L$ is required; initial impurity spin states are measured and known to be in the anti-parallel alignment. By contrast, in the $\Delta_L \neq 0$ protocol, party A (B) is subject to a Zeeman splitting in $+z$ ($-z$) direction, which has been and will be turned on at all times; the $\Delta_L$ is designed to pin the impurity spins in the desired anti-parallel state only initially, but not after the quench; therefore, we will assign small $\Delta_L$ and choose a Fermi energy $E_F$ just slightly above the ground state, in order to prevent involving other unwanted impurity spin configurations. The schematics in the insets, lower right corner in Fig. 1, exhibits the two protocols regarding impurity-electron band structures; they will serve as marks to indicate the used protocols throughout all restuls/figures herein. Note, however, that both protocols adopt unpolarized electron spins as the ICs, i.e., $\Delta_L$ is locally applied to parties A and B but *not* to C. The device mimics the FET geometry in that the gate voltages can be introduced, particularly within the spacing region, to tune the entanglement.

The time-dependent Hamiltonian for the device in Fig. 1 reads

$$H(t) = -\gamma \sum_{\langle ij \rangle \sigma} c_{i\sigma}^\dagger c_{j\sigma} - \sum_{i \in A} \frac{\Delta_L}{2} S_{i,z} + \sum_{i \in B} \frac{\Delta_L}{2} S_{i,z} + \delta H(t). \quad (1)$$

Defining $H(t=0) \equiv H_0$, the dynamics after $t = 0$ is invoked by

$$\begin{aligned} \delta H(t) &= H(t) - H_0 \\ &= \theta(t)(-J_{sd}) \sum_i \vec{\sigma}_i \cdot \vec{S}_i \\ &\quad + \theta(t - t_{g,on}) eV_g \sum_{i^*\sigma} c_{i^*\sigma}^\dagger c_{i^*\sigma}. \end{aligned} \quad (2)$$

Here $c_{i\sigma}^\dagger$ ($c_{j\sigma}$) denotes the creation (annihilation) operator creating (annihilating) an electron of spin-z, $\sigma = \uparrow_e$ for spin-up and $\sigma = \downarrow_e$ for spin down, at site $i$. $\theta(t)$ is the unit step function. Party C contains electrons with kinetic hopping $\gamma$ between two nearest sites $\langle ij \rangle$ and with spins $\vec{s}_i = \vec{\sigma}_i \hbar/2$ at site $i$. Party A (B) consists of local impurity spins $\vec{S}_i$, in unit of $\hbar$, at site $i$ with $i \in A$ ($i \in B$). The impurities or qudits can be subject to local magnetic field which induces Zeeman splitting $\Delta_L$ in $\pm z$-direction. The sd exchange coupling of strength $J_{sd}$, present at $t = 0$, introduces quench dynamics. An additional gate voltage can be applied at time $t_{g,on}$ and at sites $i^*$ to the electrons. Although the dimension of the Hilbert space is not shown explicitly in Eqs. (1) and (2), note each term lives in the space of dimension same as $I^C \otimes I^A \otimes I^B$. Here $I^C$ is the identity operator in the electron spin-1/2 Fock space, while $I^A$ ($I^B$) the identity in the spin space of dimension $(2S+1)^{N_A}$ [$(2S+1)^{N_B}$] with $N_A$ ($N_B$) being the number of impurities in party A (B). Below, the operators and wave functions all can be enlarged to match the dimension $I^C \otimes I^A \otimes I^B$ by proper tensor products with identities. However, to keep the notation simple, we will imply the dimension implicitly in the following expressions.

Before $t = 0$, the system is in a steady state characterized by the stationary wave function in the source and sample region as

$$\Psi_\alpha^{st}(E) = G^R(E) \xi_\alpha(E) \sqrt{v_\alpha}. \quad (3)$$

Here $\alpha$ denotes the modes in both the left $p = Left$ and right $p = Right$ leads of the transverse wave function $\xi_\alpha$ with group speed $v_\alpha$. The $\xi_\alpha$ and $v_\alpha$ are obtained by solving a generalized eigenvalue equation [cf. Eq. (35) in Ref. [58]] and selecting half (for example, right-propagating and/or right-decaying) of the modes. The retarded GF at energy $E$ reads $G^R(E) = [E - H_0 - \Sigma^R(E)]^{-1}$. The retarded self-energy $\Sigma^R(E) = \sum_{p=Left,Right} \Sigma_p^R(E)$, assigned in the source layers, takes into account the semi-infinite leads. Specifically, if there are totally $W$ modes, in the matrix form by defining

$$Z = \left( \sqrt{v_1} \overset{|}{\underset{|}{\xi_1}}(E), \ \sqrt{v_2} \overset{|}{\underset{|}{\xi_2}}(E), \ \cdots \ \sqrt{v_W} \overset{|}{\underset{|}{\xi_W}}(E) \right) \quad (4)$$

and the diagonal matrix, of Fermi-Dirac distribution $f_\alpha(E) = 1/\{1 + \exp[(E - E_F)/k_B T]\}$,

$$F = \begin{pmatrix} f_1(E) & & 0 \\ & \ddots & \\ 0 & & f_W(E) \end{pmatrix}, \quad (5)$$

one has finite level broadening $ZZ^\dagger = \Gamma = i[\Sigma^R - (\Sigma^R)^\dagger]$ and the lesser self-energy $\Sigma^<(E) = iZFZ^\dagger$. In other words, Eqs. (3), (4), and (5) yield the lesser GF $G^<(E)$,

$$i \sum_\alpha f_\alpha(E) \Psi_\alpha^{st}(E) [\Psi_\alpha^{st}(E)]^\dagger = G^R(E) \Sigma^<(E) G^A(E),$$

with the advanced GF $G^A(E) = [G^R(E)]^\dagger$. We note that in both of the above mentioned protocols, the $\Psi_\alpha^{st}(E)$ can be expressed in the separable form,

$$\Psi_\alpha^{st}(E) = \psi_\alpha^e(E) \otimes \chi_A \otimes \chi_B,$$

by the electron $\psi_\alpha^e(E)$ wave function in party C and the local spin $\chi_A$ ($\chi_B$) wave function in party A (B).

After $t = 0$, the quench renews the wave function with the additional dynamical wave function $\Psi^d$ that satisfies the differential equation,

$$\begin{aligned} i\hbar \frac{\partial}{\partial t} \Psi_\alpha^d(E, t) &= [H(t) + H_{M_\alpha(t)} - E] \Psi_\alpha^d(E, t) \\ &\quad + \delta H(t) \Psi_\alpha^{st}(E), \end{aligned} \quad (6)$$

of the IC, $\Psi_\alpha^d(t=0) = 0$. Thus, the dynamical wave function is invoked only when $\delta H(t)$ is present. The total wave function contributed from mode $\alpha$ can then be computed via

$$\Psi_\alpha(E,t) = \left[\Psi_\alpha^d(E,t) + \Psi_\alpha^{st}(E)\right] e^{-iEt/\hbar}. \quad (7)$$

Here $H_{M_\alpha}$ in Eq. (6) describes the buffer Hamiltonian at mode $\alpha$ in the left (if $\alpha \in Left$) and right (if $\alpha \in Right$) leads. Each buffer layer has the same Hamiltonian, and the hopping $\gamma$ between adjacent layers are also identical. The $M_\alpha(t)$ increases with time. When solving the differential Eq. (6), the $\Psi_\alpha^d$ prorogation is monitored; for a given solution at time $t$, if the last/farthest buffer contains nonzero $\Psi_\alpha^d$, then increase $M_\alpha$, return to the previous time step, and resolve the differential equation with the enlarged $M_\alpha$. One can, in practice, conveniently set $M_\alpha(t=0) = 1$ as the IC. In our simulations on Eq. (6), the Dormand-Prince method is implemented, with tolerances obtained by comparing the fifth-order solution with the fourth-order, making the step size adaptive. Meanwhile, the used buffer $M_\alpha$ is extrapolated at a future time; we find fitting with quadratic polynomials will be feasible. With Eq. (7), at a given time $t$ and energy $E$, the DM incorporated with the distribution $f_\alpha(E)$ can then be computed by

$$\varrho(E,t) = \sum_\alpha f_\alpha(E) \Psi_\alpha(E,t) \Psi_\alpha^\dagger(E,t), \quad (8)$$

and the lesser GF by,

$$G^<(E,t) = i\varrho(E,t),$$

which is used to evaluate [via Eq. (31) in Ref. [84]] electron bond currents that amount to the cross-sectional currents shown in this paper.

We discuss the behavior in the long-time limit and address how dissipations will lead to a new steady state. Despite the main focus of the paper being on the early growth of the entanglement, the limit $\Psi_\alpha(E,t \to \infty)$ provides valuable insights and references for choosing parameters to avoid numerically-unobservable small entanglement. Additionally, as demonstrated later, this limit is also relevant to the fate of entanglement. Assume first that, whatever time-dependent Hamiltonian presents in our sample, $\delta H$ eventually ends up as a time-constant $\delta H(t \to \infty) = K$. According to Eq. (6), the generic form arises, with all $M_\alpha = M \to \infty$,

$$\begin{aligned}\Psi_\alpha^d(E, t \to \infty) &\approx e^{-i[H_0 + K + H_{M \to \infty} - E]t/\hbar} X \\ &+ G_K^R(E) K \Psi_\alpha^{st}(E).\end{aligned} \quad (9)$$

Here $X$ and $K$ are generally spatially varying but time-independent matrices. If any imaginary energy $E \mapsto E + i\eta$ is assigned with finite $\eta > 0$ quantifying any dissipation processes, e.g., originating from finite quasi-particle life time, the first term in Eq. (9) then depicts the transient response since it dies out in the long-time limit. Accordingly, the new (with $K$) steady state is described by

$$\Psi_\alpha^d(E, t \to \infty) \approx G_K^R(E) K \Psi_\alpha^{st}(E). \quad (10)$$

We are interested in the sample-source region, where the new steady retarded GF $G_K^R(E) = [E - H_0 - K - H_{M \to \infty}]^{-1}$ reduces to

$$G_K^R(E) = \left[E - H_0 - K - \Sigma^R(E)\right]^{-1}, \quad (11)$$

provided that in the above Eq. (11), one sets $M_\alpha = 0$ and restores $\Sigma^R(E)$ back to the source layers, same as adopted in the wide-band-limit approximation where $H_{M_\alpha(t)}$ is replaced with $\Sigma^R(E)$ in Eq. (6). Equations (10) and (11) allow us to reach the new lesser GF as well as the DM $\varrho$ by noting

$$\begin{aligned}\Psi_\alpha(E, t \to \infty) &= \left[\Psi_\alpha^d(E, t \to \infty) + \Psi_\alpha^{st}(E)\right] e^{-iEt/\hbar} \\ &\approx \left[G_K^R(E) K + G_K^R(E) G^R(E)^{-1}\right] \\ &\quad \times \Psi_\alpha^{st}(E) e^{-iEt/\hbar} \\ &= G_K^R(E) G^R(E)^{-1} \\ &\quad \times \Psi_\alpha^{st}(E) e^{-iEt/\hbar}.\end{aligned} \quad (12)$$

For finite $\eta$ in the $t \to \infty$ limit, using Eqs. (8) and (12) we eventually arrive at

$$\begin{aligned}\varrho(E, t \to \infty) &= G_K^R(E) G^R(E)^{-1} Z F Z^\dagger \\ &\quad \times G^A(E)^{-1} G_K^A(E) \\ &= G_K^R(E) \Sigma^<(E) G_K^A(E)/i \\ &= G_K^<(E)/i,\end{aligned}$$

with the lesser GF $G_K^<(E) = G_K^R(E) \Sigma^<(E) G_K^A(E)$ at the new steady state. The corresponding $K$ for our present system can be readily identified by referring to Eq. (2). Although dissipations play an important role in determining the fate of entanglement, here except in Figs. 3 and 4, all our figures retain the transient $X$ term in Eq. (9). Therefore, the main focus of this paper is on the early-time responses, including the birth and early growth of entanglement, as well as the non-dissipative responses.

### B. Entanglement monotone

Tracing out environmental electron spin degrees of freedom in Eq. (8) yields the reduced DM

$$\rho = Tr_s(\varrho),$$

which can be used to calculate different monotones for the $A$-$B$ entanglement. *Only for pure states*, $Tr\rho^2 = 1$, the monotones addressed below are faithful, namely, nonzero monotones$\Leftrightarrow$entanglements. On the other hand, for mixed states $\rho = \sum_n w_n \rho_n$, with $Tr\rho^2 < 1$ and $Tr\rho_n^2 = 1$, as encountered presently after the quench, both classical correlations (due to the weighting constraint $\sum_n w_n = 1$) and quantum entanglements can manifest. The monotone dubbed MI

$$\mathcal{M}(\rho) = \mathcal{S}(\rho^A) + \mathcal{S}(\rho^B) - \mathcal{S}(\rho^{AB})$$

including both the classical correlations and quantum entanglements via the Von Neumann entropy $\mathcal{S}(\varrho) = -Tr(\varrho \ln \varrho)$. Here $\rho^A = Tr_{\bar{A}}(\varrho)$ stands for the reduced DM by tracing out all other degrees of freedom $\bar{A}$ not belonging to $A$. Similarly, the conventional notation $\rho = \rho^{AB}$ is used with

$$\rho^{AB} = Tr_{\bar{A}\bar{B}}(\varrho)$$

denoting the DM $\rho^{AB}$ by tracing out all other degrees of freedom $\bar{A}\bar{B}$ not belonging to $AB$, $A+B$, or $A \cup B$. As an example, the equal-weight mixing of the two pure unentangled states, $\rho = |\uparrow\uparrow\rangle \langle\uparrow\uparrow|/2 + |\downarrow\downarrow\rangle \langle\downarrow\downarrow|/2$, is still unentangled, while it renders nonzero, $\mathcal{M}(\rho) = \ln 2$, classical correlations.

In the case of two-qubit entanglement, regardless of whether the states are pure or mixed, the concurrence $\mathcal{C}(\rho)$ faithfully detects the entanglement by finding the maximum

$$\mathcal{C}(\rho) = \max(0, \lambda_1 - \lambda_2 - \lambda_3 - \lambda_4)$$

with eigenvalues $\lambda_1 \geq \lambda_2 \geq \lambda_3 \geq \lambda_4$ of the matrix $\sqrt{\sqrt{\rho}\rho'\sqrt{\rho}}$. Here $\rho'$ is constructed by the conjugate $\rho^*$ of the four-by-four DM and the square product $\sigma_y^{\otimes 2} = \sigma_y \otimes \sigma_y$ of the Pauli spin $\sigma_y$ matrix as

$$\rho' = \sigma_y^{\otimes 2} \rho^* \sigma_y^{\otimes 2}.$$

Another monotone, which also faithfully reflects the two-qubit entanglement, is the LN

$$\mathcal{L}(\rho) = \ln\left(Tr\sqrt{(\rho^{T_A})^\dagger \rho^{T_A}}\right),$$

by probing the negative eigenvalues in the partial transpose,

$$(\rho^{T_A})_{ab,a'b'} = \rho_{a'b,ab'}.$$

Here subscripts $a$ and $a'$ ($b$ and $b'$) denote the indices belonging to $A$ ($B$). Note $Tr\sqrt{(\rho^{T_A})^\dagger \rho^{T_A}} = Tr\sqrt{(\rho^{T_B})^\dagger \rho^{T_B}}$ independent of what party is transposed. Also, note that if $A$ and $B$ are unentangled even in a mixed state, then the system depicted by the partial transpose $\rho^{T_A}$ remains physical, i.e., having nonnegative eigenvalues. Nevertheless, for mixed states beyond two qubits ($SN_A > 1/2$ or $SN_B > 1/2$), the converse statement does not always hold. That is, $\mathcal{L}(\rho) > 0$ in general is a sufficient but not a necessary condition for entanglement, namely, nonzero $\mathcal{L}(\rho) > 0 \Rightarrow$ entanglement; zero $\mathcal{L}(\rho) = 0$ says nothing about the entanglement beyond two qubits in mixed states.

To locate the birth of $A$-$B$ entanglement, we introduce the environment support states (ESSs), which we conveniently label as $|C_\nu, A, B\rangle$. Here $|A, B\rangle$ represents the collective impurity spin state. The ESSs tailor the environment size $\nu$ (electron degrees of freedom in the present system) supporting the $A$-$B$ entanglement. We interpret $\nu$-resolved $\mathcal{C}(\rho_\nu)$, $\mathcal{L}(\rho_\nu)$, and $\mathcal{M}(\rho_\nu)$ as the amount, quantified by their corresponding monotones, of $A$-$B$ entanglement donated by the environmental electrons when staying at $\nu$. The normalization $\rho_\nu \mapsto \rho_\nu/Tr\rho_\nu$ is ensured so that the $\nu$-resolved monotone can then be computed. As an example, if $\nu$ comprises only one site $\nu = i$, then $\rho_i$ is the DM $\rho$ projected onto the $C_i + A + B$ subspace. This subspace is spanned by the state vectors incorporating the local-electron-site of the form $|C_i, A, B\rangle$. $\mathcal{C}(\rho_i)$ quantifies the entanglement contributed from electrons at site $i$. As shown later, the site-resolved $\mathcal{C}(\rho_i)$ and $\mathcal{L}(\rho_i)$ help grasp ESDs in the two-qubit case. Similarly, we can extend the ESSs to encompass a layer $\nu = l$. The layer-resolved $\mathcal{L}_l = \mathcal{L}(\rho_l)$ and $\mathcal{M}_l = \mathcal{M}(\rho_l)$ uncover their corresponding layers of birth. The monotones below, unless further specified, are computed via the DM projected onto the sample-source region, which we refer to as the active region. Another approach known as the hydrodynamic limit is often useful for predicting the long-time behavior. In this limit, the wave front of the entanglement or light-cone spreading of the entanglement [31] is monitored, and one increases the sample $L$ to follow the light-cone, so that $L/t$ remains finite as $t \to \infty$. Here, to study the birth and early growth, we will consider entanglement primarily supported by the active region of fixed $L$, while the entanglement growth after saturation can be predicted using a similar volume-scaling law [26, 27].

## III. RESULT AND DISCUSSION

Our findings are presented and analyzed in this section. In the following simulations, all energy parameters are in units of hopping energy $\gamma$. Time is measured in units of $\hbar/\gamma$, which corresponds to approximately 0.66 picoseconds when adopting $\gamma = 1$ meV. In the (first) $\Delta_L = 0$ protocol, we set $E_F = -\gamma$. In the alternative (second) protocol with $\Delta_L = 10^{-3}\gamma$, Fermi-energy $E_F$ is chosen to be $E_F = E_{\min} + 10^{-4}\gamma$, slightly above the ground-state energy $E_{\min}$ to avoid involving any excited energy bands. Without loss of the general characteristics of the entanglement dynamics depicted below, we will focus on contributions only from the Fermi-surface electrons at zero temperatures. Specifically, we will consider $\varrho(t) = \varrho(E_F, t)$ and $\rho(t) = \rho(E_F, t)$. As will be demonstrated, this specific focus with $E = E_F$ permits observing the electron motions leading to the formation of the $A$-$B$ entanglement. We employ $L = 100$ sample layers, each possessing a width of three atomic sites. Unless otherwise specified, the following default param-eters are used, $J_{sd} = 0.9\gamma$ for the $\Delta_L$-free protocol and $J_{sd} = 0.18\gamma$ for the protocol requiring $\Delta_L$. Spacing layers are set at $D = 6$. Note, except in Figs. 3 and 4, the transient term in Eq. (9) is included in all figures below with $\eta = 0$. Figures 3 and 4 depict the entanglement in the $J_{sd}$-renewed steady state after the transient response has completely disappeared.

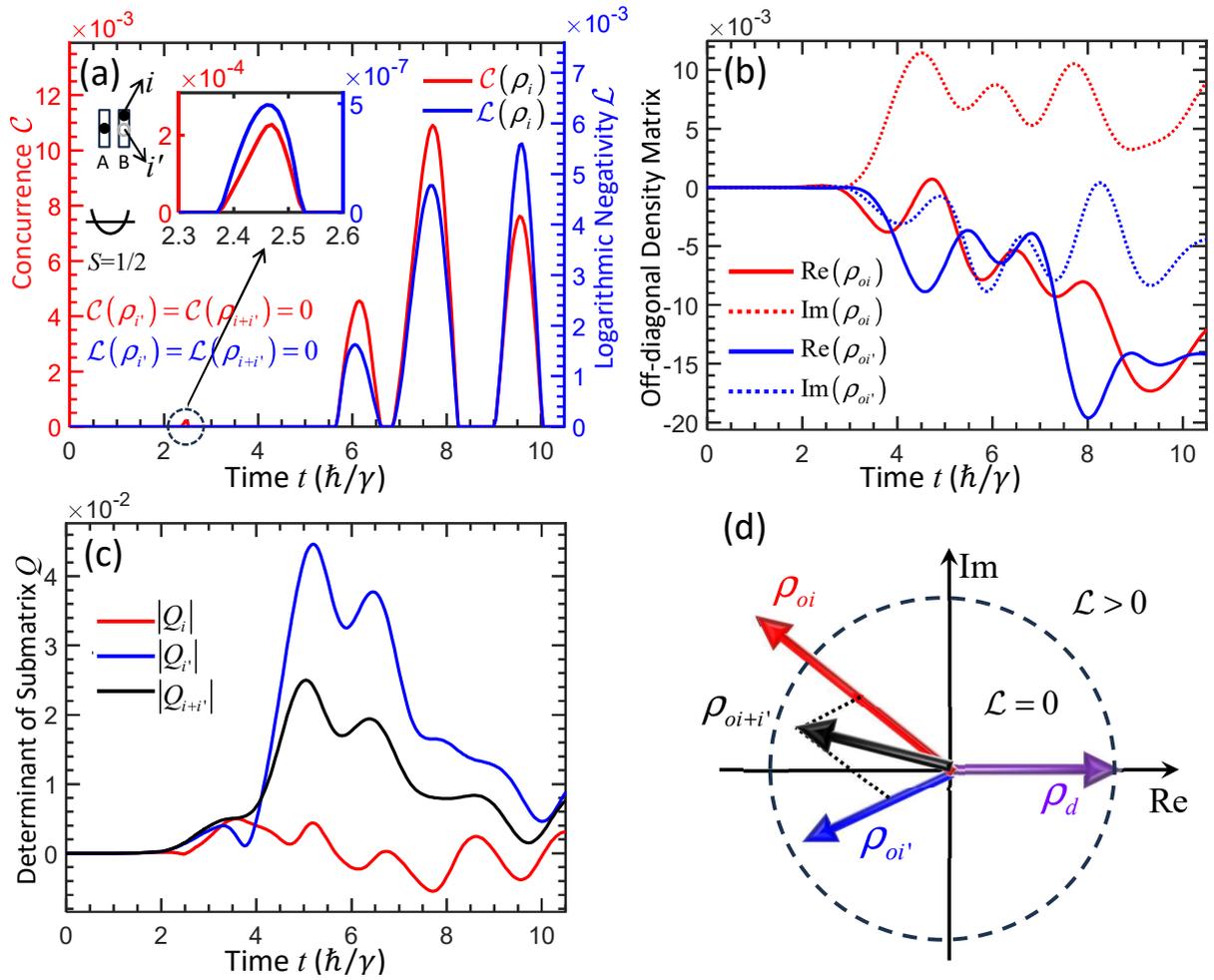

FIG. 2. (a) Two-qubit entanglement probed by site-resolved concurrence $\mathcal{C}$ and logarithmic negativity (LN) $\mathcal{L}$ as a function of time $t$. Schematics in (a) exhibits the location of the qubits in parties $A$ and $B$ distant by $D = 6$ spacing layers. It also indicates the examined environment sites $i$ and $i'$. Inset in (a) shows the zoom into $t \approx 2.45\hbar/\gamma$. (b) Real and imaginary values of the elements of the site-resolved density submatrix $Q$ defined in Eq. (15). (c) Matrix determinant of $Q$ from site $i$ ($Q_i$) and from site $i'$ ($Q_{i'}$) as well as sites $i \cup i'$ ($Q_{i+i'}$). (d) Schematics in the complex plane illustrating the diagonal $\rho_d$ and off-diagonal $\rho_o$ in Eq. (15) supported by $i$, $i'$, and both, denoted by $\rho_{oi}$, $\rho_{oi'}$, and $\rho_{oi+i'} = (\rho_{oi} + \rho_{oi'})/2$, respectively. In (d), if the length of arrow representing $\rho_o$ exceeds the dashed circle set by $\rho_d$, entanglement $\mathcal{L} > 0$ emerges.

In the following sections, we first examine the case without the gate voltage in Secs. III A and III B. We begin with the minimal set, which consists of the smallest total spin in parties $A$ and $B$. Then, we explore entanglement beyond the minimal set, where at least one of the parties, $A$ or $B$, possesses a total spin $SN_{A/B} > 1/2$ (e.g., a qutrit or more than two qubits). We then further introduce the gate voltage in Sec. III C to unveil its influence on entanglement. *Without further specification,* *both protocols share the same features described below.*

### A. Entanglement in minimal set of two qubits

Consider the minimal set containing two qubits, one qubit in each party $A$ and $B$. Before delving into our system, it is insightful to mention an example illustrating how enlarging $\nu$ can eliminate the entanglement. In this example, let $\nu$ comprise two sites, $i$ and $i'$. We assume that $i$ and $i'$, supporting $\rho_i$ and $\rho_{i'}$, are described by the two Bell states $\left|\Phi_i^+\right\rangle = (\left|\uparrow\uparrow\right\rangle + \left|\downarrow\downarrow\right\rangle)/\sqrt{2}$ and $\left|\Phi_{i'}^-\right\rangle = (\left|\uparrow\uparrow\right\rangle - \left|\downarrow\downarrow\right\rangle)/\sqrt{2}$, respectively. Here $\left|S^A, S^B\right\rangle$ indicates the qubit spin state $S^A$ in party $A$ and $S^B$ in party $B$. Both $\rho_i = \left|\Phi_i^+\right\rangle\left\langle\Phi_i^+\right|$ and $\rho_{i'} = \left|\Phi_{i'}^-\right\rangle\left\langle\Phi_{i'}^-\right|$

are entangled pure states. However, with the equal-weight mixture that includes both $i$ and $i'$ to $\nu$, one turns the two entangled states into a $\rho_{i+i'} = (\rho_i + \rho_{i'})/2 = (|\uparrow\uparrow\rangle \langle\uparrow\uparrow| + |\downarrow\downarrow\rangle \langle\downarrow\downarrow|)/2$ unentangled one. It is worth noting that this example has a small likelihood of occurring, as it necessitates a highly specific, namely equal-weight, mixture. In our two-qubit system, the suppression of entanglement is more pronounced, for it takes place over a wide range of mixtures.

Return to our present system. For $\nu$ contains the whole active region, no entanglements are found, $\mathcal{C}(\rho) = 0$ and $\mathcal{L}(\rho) = 0$. Narrowing $\nu$ down to a layer $\nu = l$ still yields zero $\mathcal{C}_l = \mathcal{C}(\rho_l) = 0$ and $\mathcal{L}_l = 0$. Shrink the ESSs until $\nu$ comprises only one site, nonzero $\mathcal{C}(\rho_i)$ and $\mathcal{L}(\rho_i)$ are eventually captured. Figure 2 shows the site-resolved quantities by choosing $i$ at the qubit location $z = 3$ in party $B$ and $i'$ in the same layer at $z = 2$ below $i$. The $\mathcal{C}(\rho_i)$ and $\mathcal{L}(\rho_i)$ in Fig. 2(a) illustrate four ESDs (three rebirths) within the chosen time window $0 \leq t \leq 10\,(\hbar/\gamma)$. The two-site supported $\mathcal{C}(\rho_{i+i'}) = 0$ and $\mathcal{L}(\rho_{i+i'}) = 0$ all vanish. Although specific distribution of the two qubits is adopted, as shown in the schematics of Fig. 2(a), the features mentioned above do not depend on the distribution. The two-qubit entanglement appears to favor support from a smaller environment, which is intriguing.

To comprehend this behavior, we analyze the negativity in the partial transpose. In the present spin-$z$ anti-parallel configuration, the spin-transfer torques by $J_{sd}$ are purely quantum [56, 57], rendering zero impurity spin-$x$ $S_x^{A/B}$ and spin-$y$ $S_y^{A/B}$ expectation values, $Tr\left(\rho S_x^{A/B}\right) = Tr\left(\rho S_y^{A/B}\right) = 0$. Hence, the time-dependent DM takes the form

$$\rho(t) = \begin{pmatrix} \rho_{d\uparrow} & 0 & 0 & 0 \\ 0 & \rho_{\uparrow\downarrow} & \rho_o^* & 0 \\ 0 & \rho_o & \rho_{\downarrow\uparrow} & 0 \\ 0 & 0 & 0 & \rho_{d\downarrow} \end{pmatrix}. \quad (13)$$

For the above matrix representation (13), the impurity spin-$z$ basis is arranged in the following order, $\left|S_z^A, S_z^B\right\rangle = |\uparrow\uparrow\rangle, |\uparrow\downarrow\rangle, |\downarrow\uparrow\rangle$, and $|\downarrow\downarrow\rangle$. For instance, the diagonal $\rho_{d\uparrow} \equiv \rho_{\uparrow\uparrow}$ is on the basis operator $|\uparrow\uparrow\rangle\langle\uparrow\uparrow|$ and off-diagonal mixed $\rho_o \equiv \rho_{\downarrow\uparrow,\uparrow\downarrow} = \rho_{\uparrow\downarrow,\downarrow\uparrow}^*$ is on the basis operator $|\downarrow\uparrow\rangle\langle\uparrow\downarrow|$. The IC enters through $\rho_{\uparrow\downarrow}$; as time progresses, we find that all the other nonzero elements in Eq. (13) undergo development of similar magnitudes.

Note that the partial transpose

$$\rho^{T_B}(t) = \begin{pmatrix} \rho_{d\uparrow} & 0 & 0 & \rho_o^* \\ 0 & \rho_{\uparrow\downarrow} & 0 & 0 \\ 0 & 0 & \rho_{\downarrow\uparrow} & 0 \\ \rho_o & 0 & 0 & \rho_{d\downarrow} \end{pmatrix} \quad (14)$$

is block-diagonal, i.e., the block submatrix

$$Q = \begin{pmatrix} \rho_{d\uparrow} & \rho_o^* \\ \rho_o & \rho_{d\downarrow} \end{pmatrix} \quad (15)$$

in Eq. (14) is decoupled from $\rho_{\uparrow\downarrow}$ and $\rho_{\downarrow\uparrow}$. Since all diagonals in matrix (14) consist of positive real numbers, any negative eigenvalues of $\rho^{T_B}$, if they exist, originate solely from $Q$. The eigenvalues of $Q$ are given by $\left(TrQ \pm \sqrt{(TrQ)^2 - 4|Q|}\right)/2$. Thus, if the matrix determinant is negative $|Q| < 0$, then there will be negative eigenvalues of $\rho^{T_B}$, which results in nonzero $\mathcal{L}(\rho) > 0$. In other words, the condition for the presence of entanglement is having larger off-diagonal elements satisfying

$$|\rho_o| > \sqrt{\rho_{d\uparrow}\rho_{d\downarrow}} \equiv |\rho_d|$$

or

$$|Q| < 0.$$

This condition by comparing the magnitude between off-diagonal $\rho_o$ and diagonal $\rho_d$ in the complex plane is schematically shown in Fig. 2(d). Note that $\rho_o$ can generally lie in any direction (of various phases), while diagonal $\rho_d$ always stays in the positive real axis (of zero phase). If we assign $\nu = i$ at the qubit location, the site-resolved off-diagonal $\rho_{oi}(t)$ can sometimes be greater than the diagonal $\rho_{di}(t)$, and at other times smaller, leading to sequential ESDs. On the other hand, if we assign $\nu = i'$ next to the qubit location, two noteworthy characteristics emerge. First, $\rho_{oi'}(t)$ is smaller than $\rho_{di'}(t)$ yielding $|Q_{i'}(t)| > 0$ in Fig. 2(c). Second, $\rho_{oi'}(t)$ is in a direction quite different from $\rho_{oi}(t)$, i.e., they are out of phase; see Figs. 2(b) and 2(d). As a result from these two characteristics, when we increase the environment degrees of freedom $\nu = i + i'$, then off-diagonal $\rho_{oi+i'} = (\rho_{oi} + \rho_{oi'})/2$ gets decreased (smeared out or averaged out), as indicated by the black arrow in Fig. 2(d). The decreasing leads to $|Q_{i+i'}(t)| > 0$ in Fig. 2(c). On the contrary, since both $\rho_{di}$ and $\rho_{di'}$ stay in the same real axis at all times, the addition $\rho_d = (\rho_{di} + \rho_{di'})/2$ in Fig. 2(d) does not encounter the same smearing. Actually, apart from the qubit site, most of the $i'$ sites exhibit $|Q_{i'}(t)| > 0$ and thus fail in building up $A$-$B$ entanglement. This elucidates why the two-qubit entanglement prefers a smaller environment $\nu$ and also provides insight into the ESDs. In line with our findings, indeed, we find that most of the existing research on ESDs pertains to the two-qubit case [49–51, 85–88]. ESDs have also been experimentally observed in the nitrogen-vacancy center in diamond [88], as well as in quantum optics experiments [51]. Motivated by the arguments above, we will revisit the two-qubit entanglement later for avoiding ESDs.

### B. Entanglement beyond minimal set: birth and early growth

We examine the system where at least one of the parties $A$ and $B$ possesses a total spin $N_{A/B}S\hbar$ greater than $\hbar/2$. In such a system beyond the minimal set, the DM $\rho$ is a matrix of dimension greater than four. Therefore, our arguments regarding ESDs in Sec. III A are not applicable, or they require modifications based on a more intricate matrix analysis. However, we find that the established entanglement beyond the minimal set is, in fact, robust; we can not identify any ESDs. Even in the presence of dissipations, the entanglement can persist in the long time $t \to \infty$.

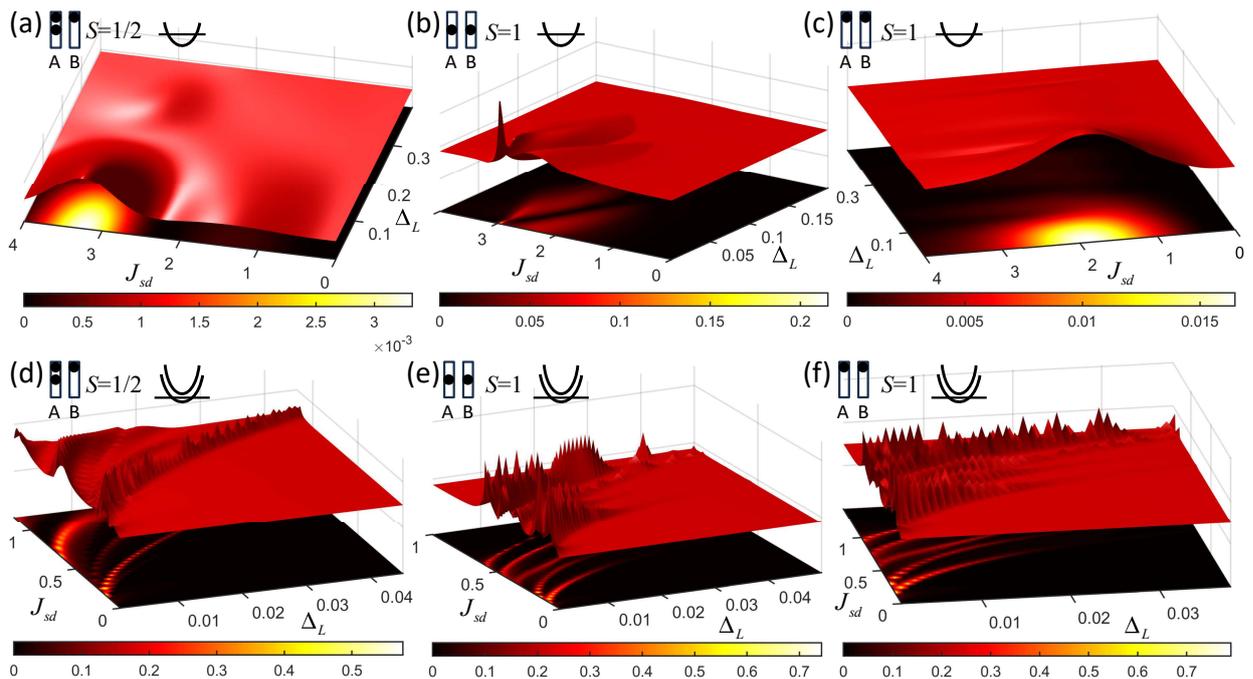

FIG. 3. Logarithmic negativity $\mathcal{L}$ as a function of $sd$ exchange $J_{sd}$ and local Zeeman splitting $\Delta_L$ with finite dissipation $\eta > 0$ in long-time $t \to \infty$ limit. In (a), (b), and (c), Fermi energy $E_F = -\gamma$ is fixed. In (d), (e), and (f), $E_F$ is adjusted according to $E_F = E_{\min} + 10^{-4}\gamma$, and a minimum $\Delta_L = 10^{-3}\gamma$ is introduced. The black dots in the schematics indicate the corresponding positions of the impurities in different parties. The $S = 1/2$ labels a qubit, and $S = 1$ a qutrit. For example, in (a) and (d), party $A$ consists of two qubits, while party $B$ of one. Parties $A$ and $B$ are separated by $D = 6$ spacing layers.

This long-time limit under dissipations of finite $\eta$ is equivalent to assigning $X = 0$ and $K = (-J_{sd}) \sum_i \vec{\sigma}_i \cdot \vec{S}_i$ in Eq. (9). With this assignment, we compute the $\mathcal{L}[\rho(t \to \infty)]$ as a function of $J_{sd}$ and $\Delta_L$ in Fig. 3. Figures 3(a), 3(b), and 3(c), show the LN in the first IC protocol with fixed Fermi energy $E_F = -\gamma$. Also, Figs. 3(d), 3(e), and 3(f), illustrate the LN in the second IC protocol with varying $E_F = E_{\min} + 10^{-4}\gamma$ slightly above the band minimum $E_{\min}$. Note that in Fig. 3, the splitting $\Delta_L$ can be applied at any finite time. Also, in Figs 3(d), 3(e), and 3(f), the minimum value $\Delta_L = 10^{-3}\gamma$ in the $\Delta_L$ axis is adopted. Since larger Zeeman splitting $\Delta_L$ implies wider energy gaps, in this case it becomes less likely for excited states to be involved in our system. As a result, the system tends to stay in the anti-parallel spin configuration that is described by our initial unentangled ground state. Conversely, $J_{sd}$ tends to drive the system towards an entangled state indicated by reducing the impurity spin-$z$ magnitudes as in the quantum spin-transfer torques [56, 57]. Indeed, Fig. 3 shows that $J_{sd} \neq 0$ is essential for having nonzero $\mathcal{L}$, i.e., all $\mathcal{L} = 0$ vanishes at $J_{sd} = 0$. We also see that for both protocols $\mathcal{L}$ is subject to an overall decay with $\Delta_L$. Furthermore, $\mathcal{L}$ also exhibits oscillatory behavior, more rapid in the second protocol than in the first. Being worth noticing, in our scan across different values of $J_{sd}$ and $\Delta_L$ (Fig. 3), the majority of the regions actually exhibit only small values of $\mathcal{L}$; larger $J_{sd}$ does not necessarily render larger $\mathcal{L}$. In the presence of dissipations, one also expects that far-distant impurities will have difficulties communicating through mediating electrons. Accordingly, one anticipates that the entanglement decays with the distance $D$ between the two parties $A$ and $B$. It is also known that a local magnetic impurity with magnetization along the $z$-axis within the bulk of an electron gas will induce a steady oscillatory spin $s_z$ distribution via the $J_{sd}$ exchange, referred to as the RKKY [59–61] oscillations. This electron spin imbalance $s_z$ diminishes when its distance to the impurity increases, weakening the exchange coupling. Hence, a decaying entanglement as $D$ increases is expected. These features are depicted in Fig. 4. The RKKY oscillations also manifest through entanglement LN, particularly when one of the impurities is situated within the bulk Fig. 4(b). Notably, after long enough separations, $D \gtrsim 40$, the information about the impurity $z$-position within each corresponding layer is lost, as if the system is a 1D chain (without width). That is, all $\mathcal{L}$ curves in Fig. 4 almost coincide after $D \gtrsim 40$.

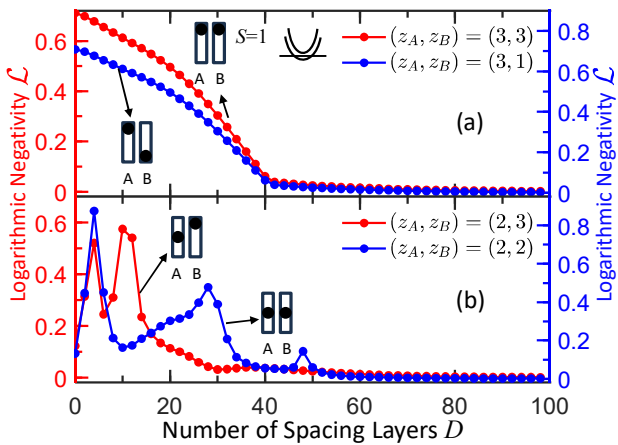

FIG. 4. Logarithmic negativity $\mathcal{L}$ as a function of number of spacing layers $D$ with finite dissipation $\eta > 0$ in $t \to \infty$ limit. The $\Delta_L = 10^{-3}\gamma$ protocol is employed. The black dots in the schematics indicate the impurity locations. In the figure legend, the $z$ positions for the impurities in parties $A$ and $B$, are also indicated by $z_A$ and $z_B$, respectively. In (a), both impurities are on the surface. In (b), at least one of the impurities is within the bulk ($z = 2$).

By restoring the transient $X$ term responsible for the non-dissipative responses in Eq. (9), we examine the birth and the early growth of the entanglement. As mentioned previously, $\mathcal{M}$ encodes both the classical and quantum correlations, while $\mathcal{L}$ only quantum entanglements. Thus, the birth of $\mathcal{M}$ will occur no later than the birth of $\mathcal{L}$. In fact, as seen in Fig. 5, $\mathcal{M}$ is born earlier than $\mathcal{L}$. Nevertheless, we can only speculate but cannot definitively conclude that the entanglement is born earlier than the classical correlation. After all, $\mathcal{L}$ serves merely as a sufficient condition for entanglement. In Fig. 5, which illustrates the development of qutrit entanglement, we also observe that both $\mathcal{L}$ and $\mathcal{M}$ take some time to appear. This reflects the finite spatial separation between parties $A$ and $B$. In other words, after the quench, it takes time for their corresponding light cones (new wavefronts) to overlap. Furthermore, one intuitive feature can be noted. When more qutrits are added to the parties, larger $\mathcal{L}$ and $\mathcal{M}$ are obtained. Also, in Fig. 5, at around $t \approx 50\hbar/\gamma$, $\mathcal{L}$ and $\mathcal{M}$ nearly saturate in the first protocol of chosen $J_{sd} = 0.9\gamma$, while they grow much more slowly in the second protocol of $J_{sd} = 0.18\gamma$.

Interestingly, shortly after the quench, $J_{sd}$ induces electron vibrating-like motions even before the presence of LN and MI. Figure 6 illustrates how electrons promptly engage in forming vibrating or oscillating currents to establish the entanglement. These cross-sectional currents vibrate with respect to the center at $x = 0$ of the spacing layers. When party $A$ has a different impurity distribution from party $B$, a net cross-sectional current can be identified through the cross-section at $x = 0$ (green lines in the middle panels of Fig. 6). To determine whether charge currents are essential for the long-term survival of entanglement, we return to Eq. (9) with assigning finite $\eta$. As $t \to \infty$, the only remaining $K$ term doesn't produce any charge currents but still yields nonzero $\mathcal{L}$ (Fig. 3). Therefore, the persistence of entanglement does not rely on charge currents. However, during the initial phase after the quench, as our results suggest, vibrating currents play a crucial role in initiating the birth of entanglement.

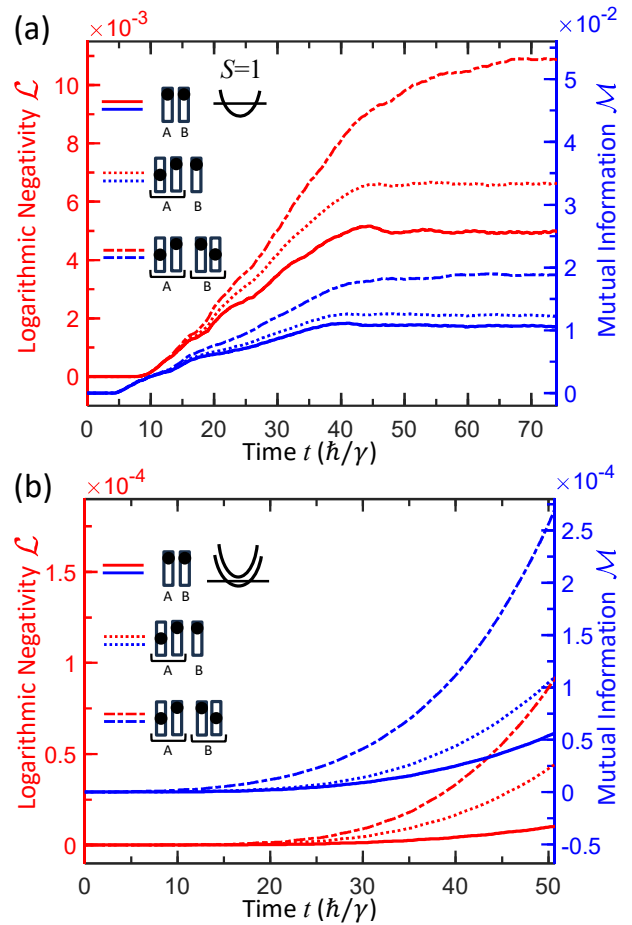

FIG. 5. Developments of logarithmic negativity $\mathcal{L}$ and mutual information $\mathcal{M}$ for qutrit entanglement in the case of (a) $\Delta_L = 0$ protocol of $J_{sd} = 0.9\gamma$ and (b) $\Delta_L = 10^{-3}\gamma$ protocol of $J_{sd} = 0.18\gamma$. The distance between parties $A$ and $B$ is $D = 6$ layers. Solid lines represent one qutrit in each party, dotted lines represent two qutrits (one) in party $A$ ($B$), and dashed lines represent two qutrits in each party. Schematics in the inset indicates the locations of the qutrits. Each qutrit is positioned at a distance of $D = 6$ layers from its nearest neighbor.

Surprisingly, the birth locations of LN and MI are quite distinct. By assigning $\nu = l$ to comprise a layer, $\mathcal{L}_l$ and $\mathcal{M}_l$ reveal their corresponding layers of origin. Figure 7 presents the snapshots captured right after the birth. Again, since $\mathcal{M}_l$ faithfully includes all, classical and quantum, correlations, and $\mathcal{L}_l$ is not a necessary condition for entanglement, namely, $\mathcal{L}_l \in \mathcal{M}_l$, we similarly observe that $\mathcal{M}_l$ shows up earlier than $\mathcal{L}_l$. Remarkably, MI is born within the spacing layers *outside* parties $A$ and $B$. On the contrary, LN is born near the edge layers, adjacent to the spacing region, hosting parties $A$ and $B$. This birth-location feature is general in that it is independent of the number of impurities and impurity distributions, as illustrated in all panels of Fig. 7. It also distinguishes LN from MI. By the same token, with $\mathcal{L}_l \in \mathcal{M}_l$, we can conjecture that classical correlations are born within the spacing layers where the communication between $A$ and $B$ takes place via mediating electrons. On the other hand, the two $\mathcal{L}_l$ in parties $A$ and $B$ are born at a distance. It is intriguing to find out, after the birth, whether the spacing layers for the communication are still a requirement for the persistence of entanglement. To answer this, we introduce the disjoint gate voltage in the next section.

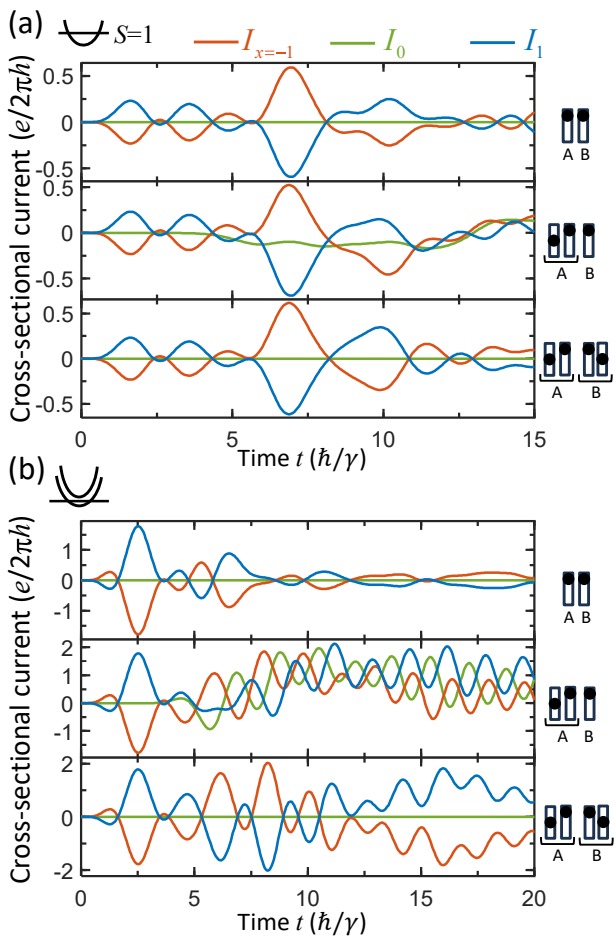

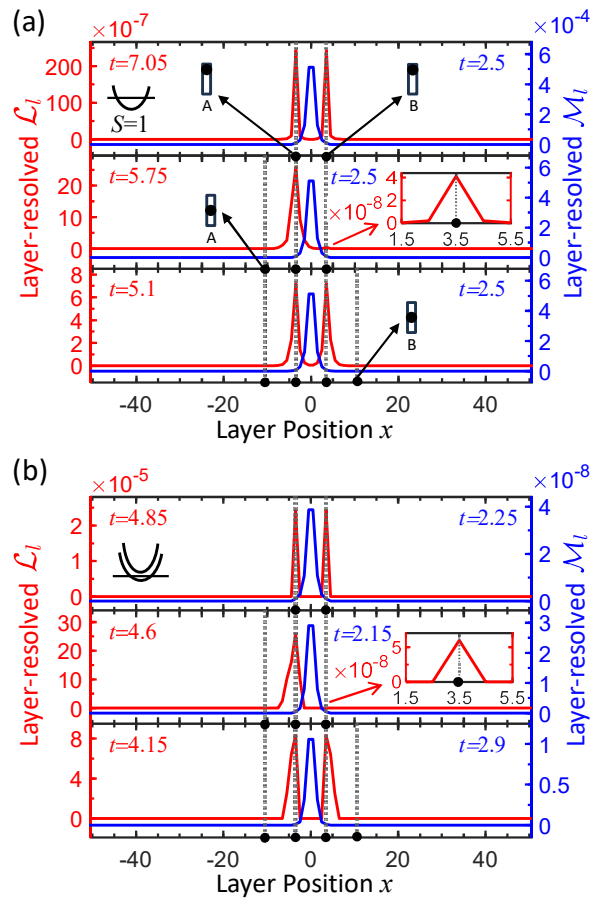

FIG. 6. With the same parameters used in Fig. 5, cross-sectional electron currents $I_x$ flowing through the plane at $x = -1$ (red lines), $x = 0$ (green lines), and $x = 1$ (blue lines), as a function of time in (a) $\Delta_L = 0$ protocol and (b) $\Delta_L = 10^{-3}\gamma$ protocol. Here $x = 0$ ($-1$ and $1$) indicates the cross-section at (next to) the spacing center, as depicted in Fig. 1. The right-flowing currents are of sign $I_x > 0$. Schematics represents the locations of the qutrits.

FIG. 7. With the same parameters used in Fig. 5, snapshots capturing the births of layer-resolved logarithmic negativity $\mathcal{L}_l$ and mutual information $\mathcal{M}_l$ in (a) $\Delta_L = 0$ protocol and (b) $\Delta_L = 10^{-3}\gamma$ protocol. The arrangement of figure panels is the same as in Fig. 6 with vertical dotted lines here further indicate the qutrit $x$ positions. The time labels in red text (for $\mathcal{L}_l$) and blue text (for $\mathcal{M}_l$) indicate the specific times in unit of $\hbar/\gamma$ when these snapshots are taken, right after the births. Insets in the middle panels show the zoom into $x \approx 3.5$, the qutrit position in $B$.

### C. Role of disjoint gate voltage

To simulate the effect that breaks down the communication between $A$ and $B$, we consider a large gate voltage $eV_g = 100\gamma$. Specifically, we impose the gate voltage within the spacing region at time $t_{g,on}$. Since our results below are virtually the same with further removing all hopping between sites adjacent and sites subject to this gating, the *quantum tunneling effects have been excluded*. In other words, $eV_g = 100\gamma$ is a gate voltage large enough to completely isolate or disjoint the gated subsystem from the rest. At the same times right after the birth as captured in Fig. 7, the gate voltages covering four layers (ranging from $x = -2$ to $x = 2$) within the spacing region are applied in Fig. 8. In Fig. 8 at $t = 80\,(\hbar/\gamma)$ long after $t_{g,on}$, we see that the layer-resolved $\mathcal{L}_l$ is entirely inhibited in the gated region, while outside this region, $\mathcal{L}_l$ develops well, exhibiting a partial suppression. Nonetheless, if the gate voltage is applied before the birth, for example $t_{g,on} = 1\,(\hbar/\gamma)$, then all negativities $\mathcal{L}_l = \mathcal{L} = 0$ are completely suppressed. This suppression is shown by the green lines in Fig. 8 as well as in Fig. 9. By applying the gate voltage at a later time $t_{g,on} = 30\,(\hbar/\gamma)$ during the growth of the entanglement, Fig. 9 shows the active-region supported $\mathcal{L}$. A similar partial suppression in the negativity is identified after $\mathcal{L}$ has reached stable saturation at some finite values.

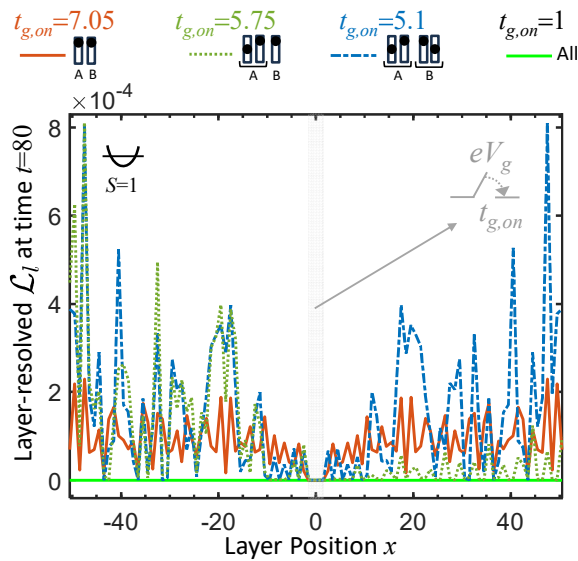

FIG. 8. Distribution of layer-resolved logarithmic negativity $\mathcal{L}_l$ in $\Delta_L = 0$ protocol, captured at time $t = 80(\hbar/\gamma)$, with disjoint gate voltage applied at time $t_{g,on}$. The corresponding qutrit locations are depicted in the top schematics of text labels indicating $t_{g,on}$, which is right after the births as identified in Fig. 7(a). Each qutrit is positioned at a distance of six layers from its nearest neighbor. The green lines represent $t_{g,on} = 1\,(\hbar/\gamma)$ for all three configurations. The gray-shaded area indicates the gating zone covering four layers within the $D = 6$ spacing region.

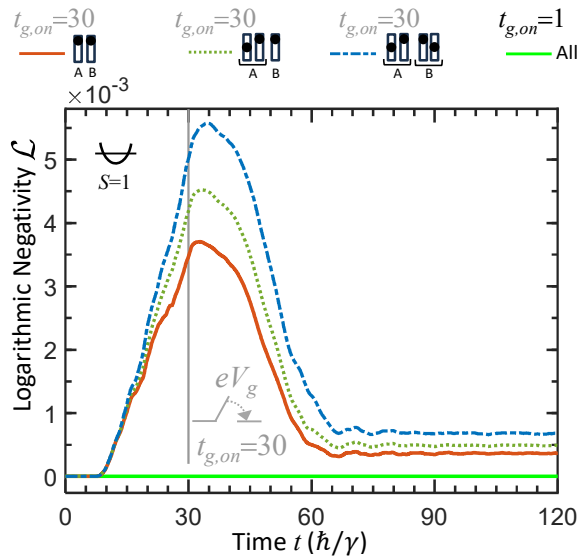

FIG. 9. Development of logarithmic negativity $\mathcal{L}$ in $\Delta_L = 0$ protocol under disjoint gate voltage applied at time $t_{g,on} = 30$ indicated by the vertical gray line. Similar to Fig. 8, top schematics represents the corresponding qutrit locations, and the same gating zone are adopted. An early applied voltage with $t_{g,on} = 1\,(\hbar/\gamma)$ is also considered, exhibited by the green lines, for all three configurations. The corresponding growth without gate voltage can be found in Fig. 5(a).

The above features are consistent with the previous findings, implying again that the charge currents are essential for forming the birth but not required in the growing phase of the entanglement. Being worth addressing, the suppression by $eV_g$ does not take place immediately after $t_{g,on}$. This delay is more obvious when there are more impurities in parties $A$ and $B$. For instance, with two qutrits in each party, as depicted by the dashed lines in Fig. 9, the suppression begins at $t \approx 35\,(\hbar/\gamma)$, which is later than $t_{g,on} = 30\,(\hbar/\gamma)$, despite the fact that the saturation time $t \approx 70\,(\hbar/\gamma)$ is nearly the same for all three configurations with one-one, two-one, and two-two qutrits in the $A$-$B$ parties. Furthermore, the saturation values of $\mathcal{L}$ actually depend on the strength of the (moderate but not disjoint) voltage $eV_g$. Thus, the system offers a gate-voltage tunable destiny of the entanglement quantified using $\mathcal{L}$.

We describe the behavior (not shown here) of MI in response to the disjoint voltage. This description also serves as an example to pinpoint classical correlations from MI. Unlike negativity, the growth of $\mathcal{M}$ remains robust against $t_{g,on}$. Independent of how early the voltage is applied, $\mathcal{M}$ grows with time. However, the layer-resolved $\mathcal{M}_l$ diminishes with sufficiently small $t_{g,on}$. For instance, if we set $t_{g,on} = 0.04\,(\hbar/\gamma)$, all $\mathcal{M}_l$ vanishes, while $\mathcal{M}$ saturates at some finite value, approximately $2.1 \times 10^{-3}$ if adopting the one-one qutrit configuration (see schematics for the red solid line in Fig. 9) of the first protocol without $\Delta_L$. This implies that in the limit $t_{g,on} \to 0^+$, the growth of $\mathcal{M}$ is attributed to the classical correlation; in other words, the resulting growth of $\mathcal{M}$ stems from the classical weighted average over unentangled pure states from each layer, characterized by $\mathcal{M}_l = 0$.

The disjoint $eV_g$ is not always destructive to entanglement. In the last part of our findings, let us revisit the case of two-qubit entanglement. We illustrate with this case that the destiny of entanglement is gate-voltage-controllable. Specifically, the disjoint voltage allows for the prevention of ESDs. Recall the analysis in Sec. III A infers that the two-qubit entanglement prefers to live in an environment of smaller degrees of freedom.

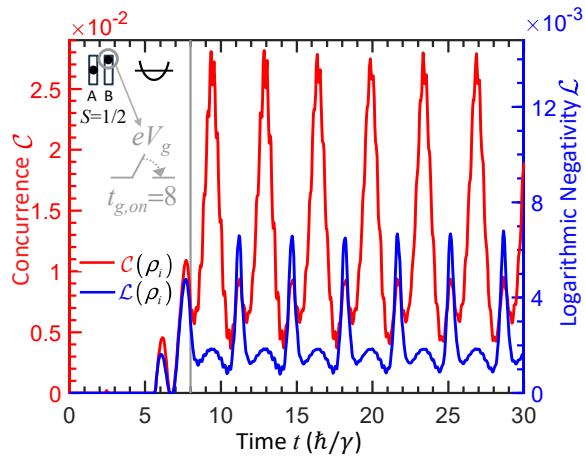

FIG. 10. Revival of entanglement from ESDs achieved by applying a disjoint gate voltage, $eV_g = 100\gamma$, to the site hosting the qubit party $B$. The voltage is turned on at time $t_{g,on} = 8\,(\hbar/\gamma)$, indicated by the gray vertical line, before the two-qubit entanglement, supported by the local site basis, dies or disappears again [refer to Fig. 2 (a)]. Note the revival can also be accomplished if $eV_g$ is applied to both of the sites hosting qubits $A$ and $B$.

Because the above large voltage is equivalent to turning off all hopping to the gated region, this shrinking of environment happens to be the role that the disjoint $eV_g$ can play. Consider $eV_g$ applied to the site that host the qubit. For example, at $t_{g,on} = 8(\hbar/\gamma)$ in Fig. 2(a) before the third complete death of the entanglement $\mathcal{C}(\rho_i) = \mathcal{L}(\rho_i) = 0$ occurs, apply the disjoint voltage to the environmental electrons at the site that couples to the $B$ qubit through $J_{sd}$. Indeed, we see in Fig. 10, the destiny of entanglement now does not involve any ESDs. With the same $t_{g,on} = 8(\hbar/\gamma)$, this prevention can also be achieved if $eV_g$ is applied to both of the sites hosting the two $A$ and $B$ qubits.

It is important to clarify the scope of our findings concerning the destiny of entanglement from the disjoint voltage. As aforementioned, $J_{sd}$-induced entanglement can sustain even in the presence of dissipation (Fig. 3). It also survives when disjoint voltage $eV_g$ is imposed (Figs. 9 and 10). Nonetheless, the *coexistence* of dissipation and disjoint voltage will turn the survivable destiny into the fatal death of entanglement. This death is predicted by the wave function in Eq. (9) of the new steady state. In Eq. (9), assigning finite $\eta$ results in solely the $K = (-J_{sd})\sum_i \vec{\sigma}_i\cdot\vec{S}_i + eV_g\sum_{i^*\sigma} c^\dagger_{i^*\sigma} c_{i^*\sigma}$ term at $t \to \infty$. Accordingly, the new steady state of the system, akin to resetting the differential equation (6) with this steady state as the new IC, is now described by two independent subsystems. One subsystem containing party $A$ and the other containing party $B$ are disjointed by $eV_g$, leading to unentanglement between $|C_A, A\rangle$ and $|C_B, B\rangle$. Here $|C_A\rangle$ and $|C_B\rangle$ represent the electron states interacting with $A$ and $B$, respectively. In other words, the herein identified tunability of the destiny through the gate voltage is only applicable within the non-dissipative regime.

## IV. SUMMARY

We summarize what has been fulfilled and discovered in this paper. By implementing the time-dependent formalism based on the wave-function dynamics satisfying differential Eq. (6), the reduced DM is constructed to investigate the bipartite $A$-$B$ entanglement resulting from the quench at $t = 0$ by the $sd$ exchange $J_{sd}$. The parties $A$ and $B$ comprise local impurities (or qudits) of opposite spin directions hosting by electrons in the open system of FET geometry. Two protocols are proposed to prepare the initial separable or unentangled states. However, most of the features found in our results are protocol-independent. We provide the long-time solution, Eq. (9), of the wave function and demonstrate how it enables us to reach the expected steady state and determine the destiny of entanglement. We mainly focus on (i) the birth and early growth of entanglement, by introducing the ESSs to quantify the supported environment of finite size $\nu$, as well as (ii) the non-dissipative $\eta = 0$ responses, by including the $X$ term in Eq. (9). In particular, the two-qubit entanglement $\mathcal{L}(\rho_i)$ in the minimal set exists with ESDs when $\nu$ contains a site $i$. The inclusion of a few sites or layers in $\nu$ leads to the disappearance of entanglement, as explained by the visualization picture in Fig. 2(d). Figure 2(d), depicting ESDs, also suggests that two-qubit entanglement prefers an environment of small $\nu$. Beyond the minimal set, the entanglement monotone $\mathcal{L}$ remains robust against ESDs. Even when dissipation $\eta > 0$ is present, it is possible to find a destiny of nonzero $\mathcal{L}(t \to \infty) > 0$ with proper choices of $J_{sd}$ and local Zeeman splitting $\Delta_L$, as shown in Fig 3. Furthermore, the entanglement exhibits an overall decay with the distance $D$ between $A$ and $B$.

The vibrating electron currents with respect to the spacing center play an essential role in fostering the birth of entanglement. However, these currents are not required to grow or maintain entanglement. The birth of MI $\mathcal{M}$ takes place within the spacing layers. By contrast, the entanglement negativity $\mathcal{L}$ is born near both of the edge layers that host parties $A$ and $B$, adjacent to the spacing region. The birth of $\mathcal{M}$ occurs earlier than $\mathcal{L}$. A large gate voltage $eV_g$ is introduced to disjoint $A$ and $B$ at time $t_{g,on}$. In the non-dissipative regime, when $eV_g$ is applied within the spacing region, partial suppression of the entanglement negativity $\mathcal{L}$ is observed. Particularly, when the voltage is applied right after the birth, the gated layers do not grow $\mathcal{L}_l$. The destiny of $\mathcal{L}(t \to \infty)$ is gate-voltage-controllable if a moderate, instead of a disjoint, $eV_g$ is applied in non-dissipative systems. The suppression takes some time to respond to the disjoint voltage, especially when parties $A$ and $B$ possess more impurities. In the case of the minimal set, the disjoint voltage applied to the site(s) hosting the qubit(s) can avoid ESDs, if $t_{g,on}$ is chosen at a time when $\mathcal{L}(\rho_i) > 0$ still survives. However, the combination of dissipation $\eta$ and the disjoint voltage $eV_g$ alters the destiny to the fate of entanglement death. In all the presented figures, we use qubits ($d = 2$) and qutrits ($d = 3$) to illustrate the above features. Nonetheless, these features are generally applicable to qudits of $d > 3$, as confirmed through several of our numerical simulations.

Although, the site-resolved $\mathcal{L}(\rho_i)$ in Fig. 10 with disjoint voltage eventually avoids encountering ESDs, and we additionally find that the layer-resolved $\mathcal{L}_l$ similarly does not encounter ESDs, the active-region-supported LN $\mathcal{L} = 0$ unfortunately remains dead. Specifically, $\mathcal{L}$ survives only in a small $\nu$ containing a few layers. Figure 2(d), with now letting $i = l$ and $i' = l'$ to be two layers, suggests that different layers are of quite different phases of $\rho_o$, leading the absence of LN. However, we regard the search for sustainable $\mathcal{L} > 0$ in a larger $\nu$ to be beyond the scope of this paper. An elaborated examination of the behavior of $\rho_o$ to discover this $\mathcal{L}$ is considered an extension of this work. In particular, beyond the antiparallel spin configurations, the present physical picture needs to be generalized.

However, the introduced size of environment $\nu$ herein would provide a useful framework for analyzing and gaining insights into entanglement. The identified characteristics of the birth and growth of negativity and mutual information contribute to our understanding of the dynamics of both classical and quantum correlations. Furthermore, the vibrating currents resulting from the quench by $J_{sd}$ sharpen our comprehension of the mechanisms responsible for entanglement birth. The revealed controllability of entanglement via gate voltage offers guidance in terms of designing solid-state-based quantum computations. We believe the findings presented here will inspire further research in the development of quantum technologies by integrating entanglement into well-established FETs.

### ACKNOWLEDGMENTS

The author thanks Wei-Chen Chien, Shih-Jye Sun, Ming-Chien Hsu, Seng-Ghee Tan, and Ching-Ray Chang for valuable discussions.